\documentclass[12pt]{iopart}
\usepackage{graphicx}
\usepackage{epsfig}
\newcommand{\nn}{\nonumber}
\newcommand{\beq}{\begin{equation}}
\newcommand{\eeq}{\end{equation}}
\newcommand{\be}{\begin{eqnarray}}
\newcommand{\ee}{\end{eqnarray}}

\def\+{\dagger}

\begin{document}
\title{ Diffuse 
cosmic gamma-rays   at 1-20 MeV: \\~A trace of the dark matter?  }
\author{  Kyle Lawson and Ariel R.Zhitnitsky}

\address{ Department of Physics and Astronomy, University of British 
Columbia, Vancouver, BC, V6T 1Z1, CANADA}

\begin{abstract}
Several independent observations of the galactic core suggest
  hitherto unexplained sources of energy. The most well known case is the 511 keV line
  which has proven very  difficult to explain with conventional astrophysical
  positron  sources. A similar, but less well known mystery is 
the excess of gamma-ray photons detected by COMPTEL across a broad energy range
$\sim 1-20$ MeV. Such photons are found to be very difficult to produce via
known astrophysical sources. We show in this work that dark matter in
  the form of dense antimatter droplets  provides  a natural 
   explanations  for the observed flux of  gamma-rays in the $\sim 1-20~$ MeV range.
   We argue that  such photons must always accompany the 511 keV line as they 
   are produced by the same mechanism within our framework.  
   We calculate the spectrum and intensity of the $\sim 1-20~$ MeV  gamma-rays,
    and find it to be consistent with the COMPTEL data.

\end{abstract}
\maketitle

\section{Introduction}
Recent observations of the galactic centre have presented a number of puzzles 
for our current understanding of galactic structure and astrophysical 
processes. In particular a series of independent observations 
have detected an excess flux of  photons across a broad 
range of energies. In particular, these observations include: 

${\bullet}$ SPI/INTEGRAL observations of the galactic centre have 
detected an excess of 511 keV gamma rays resulting 
from low momentum electron-positron annihilations. The observed intensity 
  is a mystery.    After accounting for known positron
sources, only a small fraction of the emission may be
explained~\cite{Knodlseder:2003sv,Jean:2003ci,Boehm:2003bt,
Beacom:2005qv,Zhang:2006fr,Yuksel:2006fj}.  

${\bullet}$ Detection by the CHANDRA satellite of diffuse X-ray emission 
from across the galactic bulge provides a puzzling picture:  
 after subtracting known X-ray sources  
 one finds a residual diffuse thermal X-ray emission consistent 
 with  a two-temperature plasma with 
 the hot component close to  $T\simeq 8 ~keV$. According to  \cite{Muno:2004bs}
 the  hot component 
 is very difficult to understand within the standard picture. Such a plasma
would be too hot to be bound to the galactic center. The authors of ref. \cite{Muno:2004bs} 
also remark that the energy required to sustain a  plasma of  this temperature  
corresponds to the entire kinetic energy of one supernova every 3000 yr, 
which is unreasonably high.  

\begin{figure}[t]
\begin{center}
\includegraphics[width = 0.8\textwidth]{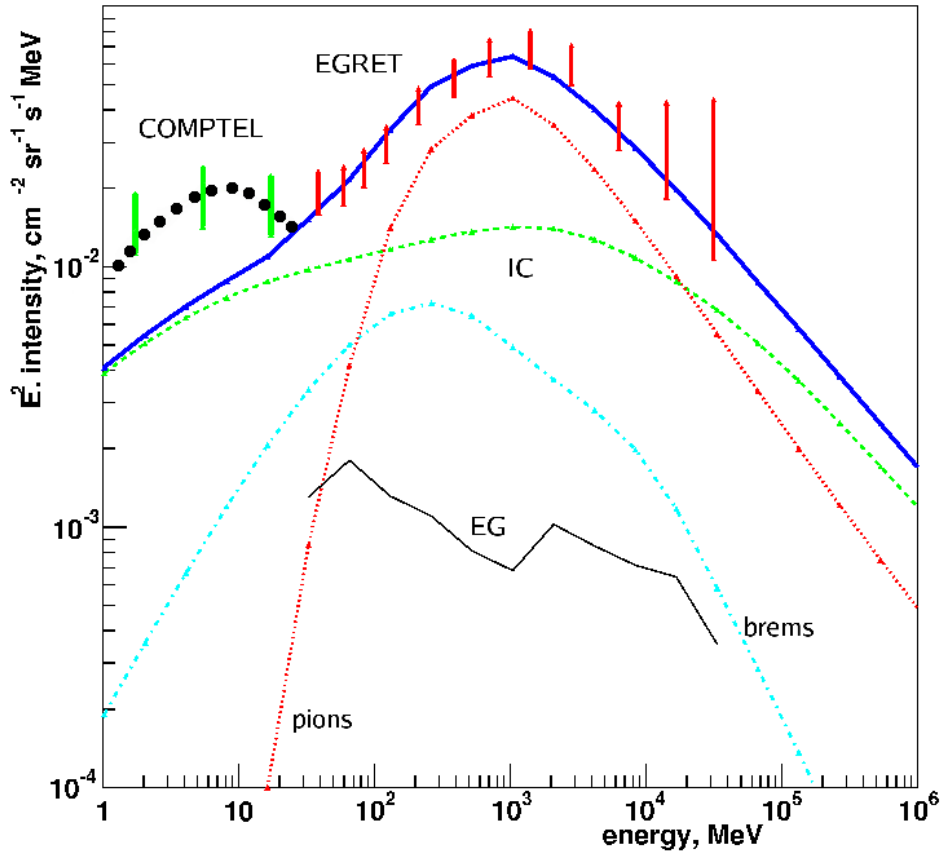}
\caption{ $\gamma$ ray spectrum of inner galaxy for optimized model\cite{Strong1}. Green vertical bars: COMPTEL data. Heavy solid line: total calculated flux for optimized  model. Heavy black dots: Combination of calculated emission spectrum from electron-nugget annihilation processes with the optimized model of \cite{Strong1}.}
 \label{spectra}
 \end{center}
 \end{figure}

${\bullet}$ The flux of gamma rays in the 1-20~ MeV range measured by COMPTEL represents yet
another mystery. As discussed in \cite{Strong} the best fit models for diffuse galactic 
$\gamma$ rays  fit the observed spectrum well for a very  broad range of 
energies, 20 MeV- 100 GeV. It also gives a good representation of the latitude 
distribution of the emission from the plane to the poles, and
 of the longitudinal distribution. However, the model fails to explain 
 the excess in the 1-20 MeV range
 observed by COMPTEL in the inner part of the galaxy ($l=330^0-30^0, |b| =0^0-5^0$),
  see figure \ref{spectra}. As claimed in ~\cite{Strong} some additional  $\gamma$ ray sources are required 
  to explain this energy region.  

These data, when taken together, suggest the existence 
of an energy source beyond currently established astrophysical 
phenomenon. The main goal of this paper is to argue that
these (seemingly unrelated) observations may be explained by a single mechanism. 
The origin of both the 511~keV radiation and broad  1 MeV $\leq k\leq $ 20~  MeV 
emission  can be naturally explained by the idea that dark
matter (DM) consists of Compact Composite Objects
(CCOs) made of matter and antimatter~\cite{Zhitnitsky:2002qa,Oaknin:2003uv,Zhitnitsky:2006vt},
similar to Witten's dense strangelets~\cite{Witten:1984rs}.   
 Dark antimatter nuggets would provide an unlimited source of positrons (e$^{+}$)  
 within this framework as
  suggested in~\cite{Oaknin:2004mn,Zhitnitsky:2006tu}. The resonance formation of positronium 
  between  impinging galactic electrons (e$^{-}$)  and positrons (e$^{+}$) from the DM nuggets, 
      and their subsequent decay, lead to the 511 keV line. Non-resonance direct 
      $e^+e^-\rightarrow 2\gamma$ annihilation   would produce a broad spectrum at 1 MeV  
$\leq k\leq $ 20~  MeV which we identify with the excess  observed by COMPTEL.
 This continuum  emission  must always accompany the  511 keV line and the two must
be spatially correlated, as argued earlier \cite{Zhitnitsky:2006vt}. 
Available observational data suggest that the intensity 
of the 511 keV line is concentrated 
in the bulge of the galaxy  (80\% of the photons come from a circle of a half-angle 6$^o$).
The {\bf excess } flux in the 1-20 MeV range observed by COMPTEL is also detected only within 
 the inner part of the galaxy ($l=330^0-30^0, |b| =0^0-5^0$). Indeed the authors of \cite{Strong}
 make the point that they have used COMPTEL measurements only "for the inner Galaxy spectra,
 since the skymaps do not show significant diffuse  emission elsewhere." 
 These observations are consistent with our proposal that both phenomena have common origin.

We stress here that   COMPTEL does indeed measure significant emissions across 
 the 1-20 MeV range from throughout the galactic disc. However the majority of this radiation
 may be attributed to well known astrophysical processes, see\cite{Strong}.
   It is only once the contributions 
 from processes such as pion decays, and inverse Compton or bremsstrahlung scattering
 of cosmic rays have been subtracted that the distribution of 
 the excess emissions from the galactic 
 centre become apparent. Whereas we assume that the 511 keV line is dominated by annihilation 
 events involving a dark matter nugget in the case of ~MeV emission a detailed subtraction of
 all contributions to the observed emissions is critical. Without such a subtraction
 the morphologies of the 511 keV line and the broad MeV emissions appear quite dissimilar.
 While present data are insufficient to determine the exact distribution of the diffuse emission
 it is found to be concentrated within the inner galaxy and our model is not in contradiction
with observations at their current resolution.
 Both further modeling of the 
 contributing processes and more detailed observations in the two different bands
 will be required to confirm or rule out 
 the validity of our proposal that the two emissions originate from the same physics.

 It has been argued that in the soft energy regime, below a few 100s of keV, total emissions are 
 dominated by the point source contribution which provides a good match to observations (with the
 exception of the 511 keV line and associated positronium continuum\cite{Lebrun}.)
 On the other hand at large energies, above 100 MeV, interstellar processes dominate and again the 
 observational data is well matched by theoretical models\cite{Strong}.
  The COMPTEL data in the 1-20MeV range 
 falls between there two separate regimes. It has proven difficult to explain by any conventional
 process.
 This is however precisely the energy range in 
 which our proposed dark matter model has observational consequences.
 
In the present paper we  estimate the intensity and the photon spectrum  in the
1 MeV   $\leq k\leq $ 20~  MeV energy range. Assuming dark antimatter to 
be the common source for the 511 keV line as well as the 1 MeV   $\leq k\leq $ 20~  
MeV emission we extract some phenomenological parameters describing their
properties. 

It is quite remarkable that another (also naively unrelated) puzzle, 
the  diffuse X-ray emission 
observed by CHANDRA~\cite{Muno:2004bs},
 may also have a common origin with the 511 keV line and excess MeV radiation 
as argued in~\cite{FZ}. We show below that dark matter in form of 
CCOs is consistent with all of
the COMPTEL data, and that it may in fact fully explain the missing sources of
emission.  If our proposal turns out to be correct,  there should be  spatial
correlations between the various emissions (511 keV line measured 
by SPI/INTEGRAL and   1 MeV  
$\leq k\leq $ 20~  MeV measured by  COMPTEL  as well as  the  diffuse 
X-ray emission with $T\sim 8$ keV measured by  CHANDRA).  
This should allow our proposal to be
verified or ruled out by future, more precise measurements.

\section{ Dark Matter as Compact Composite Objects (CCOs).}
Unlike conventional dark matter candidates, dark matter/antimatter nuggets
are strongly interacting,  macroscopically large objects.  Such a seemingly counterintuitive 
proposal does  not contradict any of the many known observational constraints on
dark matter or antimatter in our universe for three main reasons: 
  1) the nuggets carry a huge (anti)baryon charge $|B| \approx
10^{20}$ -- $10^{33}$, so they have a macroscopic size and  a tiny number density.  2) They
have   nuclear densities in the bulk, so their interaction cross-section per unit mass is small
$\sigma/M \approx 10^{-13}$ -- $10^{-9}$~cm$^2$/g. This small factor effectively
replaces a  condition on weakness of  interaction of  conventional dark matter candidates such as 
WIMPs.
3) They have a large binding energy (gap
$\Delta \approx 100$~MeV) such that baryons in the nuggets are not
available to participate in big bang nucleosynthesis (BBN) at $T
\approx 1$~MeV\@.  On large scales, the CCOs are sufficiently dilute
that they behave as standard collisionless cold dark matter (CCDM)\@.
However, when the number densities of both dark and visible matter
become sufficiently high, dark-antimatter--visible-matter collisions
may release significant radiation and energy. This obviously alters the standard 
prediction of CCDM on galactic scales. Hopefully, this radiation
can be detected and identified which would provide strong evidence 
for ``non-baryonic DM"  which nevertheless carry  huge baryon charge in the form of dense nuggets. 

The basic proposal was originally intended to explain the order 
of magnitude similarity in the energy densities of visible and dark matter,
 $\Omega_{DM} \approx 5\Omega_{B}$. 
Such a similarity has no simple explanation if dark matter represents a 
fundamentally different field than normal matter. If however, both 
visible and dark matter have their origins at the QCD phase transition 
of the early universe they would naturally be expected to have similar 
scales~\cite{Oaknin:2003uv}.  

This paper seeks specifically to explain the COMPTEL 
detection of excess photons in the 1-20 MeV range and relate this excess
to other radiation  puzzles mentioned above.  In particular, as both
resonance positronium formation and direct $e^+e^-\rightarrow 2\gamma$
photon production have a common source, the general normalization 
which depends on  DM and visible matter distributions does not bring any additional
uncertainties in our estimates which follow.

 In this paper we adopt  a simple model for nugget structure in 
 which all  quarks form one of the  color superconducting (CS) phases
  with densities a few times typical nuclear density, while  the electrons  
  in the bulk of the nuggets
   can be treated as a noninteracting Fermi gas with 
   density $n_e\simeq \frac{(\mu^2-m_e^2)^{3/2}}{3\pi^2}$,
  with $\mu$ being the electron chemical potential. A precise numerical estimation of 
$\mu$  depends on the specific details  of the CS phase under consideration,
and on the structure of the surface of the nuggets.
 Generally it may take physical values  from a few MeV up to tens of 
 MeV~ \cite{olinto}, \cite{Madsen:2001fu}. In this paper we will treat
 $\mu$ as an effective parameter of our model which varies  in the range
  MeV $ \leq \mu\leq $ tens  MeV.  Some features of the nuggets which will be relevant for
  our calculations are discussed in Appendix.

  \section{Spectrum Calculation }
As suggested  previously in refs  \cite{Oaknin:2004mn},  \cite{Zhitnitsky:2006tu} 
the 511 keV line can be naturally explained as a result of positronium formation
when a non-relativistic 
electron ($e^-$) hits the antimatter nugget surrounded by positrons ($e^+$)
 with chemical potential $\mu$. 
A certain fraction  
of galactic electrons incident on the 
CCO will annihilate directly $e^+e^-\rightarrow 2\gamma$ (rather than 
participate in  resonance positronium formation) 
resulting in the creation of photons of 
energy greater than 511keV with a maximum energy up to $\simeq \mu$.
The corresponding  fraction of electrons 
obviously depends on lepton's chemical potential $\mu$ and on specific properties
of the nugget's surface structure and the resulting distribution of positrons. 
  It is not the goal of the present paper   to calculate
the corresponding  fraction based on a specific model.
Instead, we shall   introduce this ratio   as theoretically unknown  phenomenological parameter
which will be fixed to match the observational data.
 
 It will be demonstrated that radiation arising from this DM model can  account for 
the broad spectrum across the 1-20~ MeV range observed by COMPTEL. 
Anticipating this conclusion, we should mention here, that it is quite remarkable
that the maximum energy where the excess has been observed by 
COMPTEL ($\sim $ 20~ MeV) coincides 
with a typical  estimation for  $\mu$ in quark matter \cite{olinto}, \cite{Madsen:2001fu}.
As we shall see below,   the maximum photon energy   
 within our mechanism  exactly coincides with the lepton chemical potential in the nugget
 $\mu\sim $ 20 MeV. We stress here that this energy scale arises naturally from the 
 properties of quark matter and has not been introduced in order to fit with the 
 COMPTEL observations.  This is very robust prediction of our mechanism which 
 is not sensitive to the specific details of the nugget's structure 
 nor to DM and visible matter distributions.

 As mentioned above,   we   treat the positrons at the CCO surface as 
a non-interacting Fermi gas of chemical potential $\mu $. The density 
of states in the momentum range $p$ to $p+dp$ is then given by,
$d n(p)=\frac{2d^3p}{(2\pi)^3}$
so that the rate of direct electron-positron annihilation resulting 
in a photon of momentum $k$ and  involving a 
  positron of momentum $p$ is given by, 
\beq
\frac{dI(k,\mu)}{dk dt} = \int dn(p) v(p) \frac{d\sigma (p,k)}{dk} = 
\int \frac{2d^3(p)}{(2\pi)^3}\frac{p}{E}\frac{d\sigma(p,k)}{dk},
 \eeq
where $\frac{d\sigma (p,k)}{dk} $ is the electron rest frame 
cross section for direct electron-positron annihilation resulting 
in a photon of momentum k. In this formula we assume that incoming electron has 
 a velocity $v_e$ well below that of a typical positron ($v(p)$)
 Within the nugget. Therefore,  calculations are carried out in the rest frame
 of the incident electron, i.e. $v_e=0$. 
 
The required cross-section may be obtained
 from a simple QED calculation, at tree level it is given by 
 \be
\frac{d\sigma (p)}{dk} &=&\frac{\pi\alpha^2}{mp^2}  
\left[ \frac{-(3m+E)(m+E) }{(m+E-k)^2} -2 \right]   \nn \\
&+&\frac{\pi\alpha^2}{mp^2}   \left[ \frac{ \frac{1}{k} (3m+E)(m+E)^2-
(\frac{m}{k})^2(m+E)^2}{(m+E-k)^2}\right],
 \ee
see for example\cite{Beacom:2005qv}. In this expression $E$ represents 
the energy of a Fermi gas positron.
The net production rate of photons of momentum k is then given by integrating 
this expression over all allowed momentum states of the Fermi gas
 \be
\frac{dI(k,\mu)}{dk dt} &=&\int \frac{8\pi dE}{(2\pi)^2}\frac{\pi\alpha^2}{m} 
\left[ \frac{-(3m+E)(m+E) }{(m+E-k)^2} -2\right]     \nn \\
&+&   \int \frac{8\pi dE}{(2\pi)^2}\frac{\pi\alpha^2}{m} \left[ \frac{ \frac{1}{k} (3m+E)(m+E)^2-
(\frac{m}{k})^2(m+E)^2}{(m+E-k)^2}\right],
 \ee
where we have integrated over all solid angles. In deriving this expression we
have taken into account that   the upper and 
lower limits of integration are set by the chemical potential and 
the threshold for photon production respectively. As maximum photon 
energy occurs for emission along the direction of initial positron 
momentum the threshold momentum is the value for which such a 
configuration results in a photon of energy k. Evaluating this expression 
gives the probability of annihilation  per unit time $dt$ for emitted photon energies from $k$ to $k+dk$
when a single electron hits the nugget, 
\be
\label{spectrum}
 \frac{dI(k,\mu)}{d kdt}=\frac{\alpha^2}{\pi mk^2} \Bigg [k(k^2+2mk-2m^2)\ln\bigg(\frac{(2k-m)(\mu +m-k)}{mk}\bigg) \\
 - \frac{3}{2}k^3 - (\mu+5m)k^2 + (\frac{1}{2}\mu^2 +3\mu m + \frac{9}{2}m^2)k - m^2(\mu+m) \nn \\
 +\frac{k^2 m^2}{\mu +m-k} + (8k^4-8mk^3-\frac{5}{2}m^2k^2+4m^3k-m^4)\frac{k}{(2k-m)^2}\Bigg] \nn
\ee
In the next section, based on this spectrum, we will evaluate the expected 
flux of photons in the 1-20 MeV range resulting from
from the direct  $e^+e^-\rightarrow 2\gamma$ annihilation by normalizing 
the corresponding flux to 511 keV line measured by INTEGRAL.
For normalization purposes in what follows we also need the total flux integrated over all
photon energies. For large $\mu\gg m$ the corresponding expression is given by 
\beq
\label{normalization}
 \frac{dI(\mu)}{dt}=\int  \frac{dI(k,\mu)}{d kdt} dk\simeq \frac{\alpha^2 m}{2\pi}\left(\frac{\mu}{m}\right)^{2}
 \ln\left(\frac{\mu}{m}\right) 
\eeq
This approximation is found to be fully adequate for our present purposes.

\section{ Normalization to 511 keV Line }
The original proposal \cite{Oaknin:2004mn} on formation of the 511 keV line 
assumes that almost all galactic  electrons incident on the DM anti-nuggets 
will form an intermediate state positronium. About a quarter of the positronium
decays (from the $^1S_0$ state ) release back-to-back
511~keV photons, while three quarters (from the $^3S_1$ state )
will lead to continuum emission with energy $k\leq $ 511 keV, also observed by the INTEGRAL.
 In addition, as originally suggested in
\cite{Zhitnitsky:2006vt} and as was mentioned above, 
a certain fraction of galactic electrons incident on the 
anti nuggets  will annihilate directly $e^+e^-\rightarrow 2\gamma$ avoiding 
resonance positronium formation. This direct annihilation results 
 in the creation of photons   with a maximum energy up to $\simeq \mu$
 which, by definition, is   the maximum   energy of positrons  in the nuggets.
 The corresponding spectrum for a single event was calculated above
 and is given by eq.(\ref{spectrum}). Our goal here is to present the corresponding
 expression for the  flux accounting for all annihilation processes
  happening along the line of sight towards the galactic centre.
  This flux will depend on the number of electrons along the line of sight
 which is roughly determined by  the number density of
protons, $n_{e^-}\simeq n_B\sim \rho_B(r)/m_p$ while the number density of
DM particles is determined by the DM distribution, $ n_{DM}\sim \rho_{DM}(r)(Bm_p)$.

\subsection{ Spectral flux in 1-20 MeV range from the galactic center.}
By comparing the flux for the 511 keV line with the flux in the
1-20 MeV range, one may remove
the dependence on the dark and visible matter distributions because, provided they have a 
common origin,  the radiation for both fluxes should be integrated 
along almost the same line of sight from the earth to the core of the galaxy. 
As a result, direct comparisons between the data
provides non-trivial insight into the properties of the nuggets,
independent of the matter distributions.
 Therefore,  we can   avoid the corresponding 
uncertainties related to $ \rho_{DM}(r),~~  \rho_B(r) $ as well as uncertainties related to typical
sizes of the nuggets,  their size-distribution  etc by 
  normalizing  the spectrum of these 1-20 MeV photons using  
  the well-measured intensity of the
511 keV line with an average flux observed to be $\frac{d\Phi}{d\Omega}\simeq 0.025 $ 
photons cm$^{-2}$s$^{-1}$sr$^{-1}$ coming from a circle of half angle 6$^0$
\cite{Jean:2003ci} from the inner part of the galaxy.
This region strongly overlaps with  the region of interests where COMPTEL data  
are available, ($l=330^0-30^0, |b| =0^0-5^0$). 
In what follows we neglect any differences resulting from  the 
 slightly different lines of sight for measurements by  INTEGRAL and COMPTEL.
 
In addition, we introduce the coefficient  $\chi$  as the ratio of electrons which experience
 direct $e^+e^-\rightarrow 2\gamma$ annihilation in comparison with the number 
 of electrons which experience resonance positronium formation.
Now the spectrum obtained in (\ref{spectrum}) may be normalized using the
high energy approximation (\ref{normalization}) and then scaled by the
observed flux of 511 keV photons (A more extensive description of this
flux calculation may be found in \cite{Oaknin:2004mn}). Following this procedure 
one arrives at an expression for the 
flux of 1-20 MeV photons from the bulge of the galaxy
normalized  to the 511 keV line as described above, 
\be
\label{flux}
 \frac{d\Phi(k)}{d\Omega d k}= 0.025\cdot \frac{4 \chi }{\rm MeV\cdot s\cdot cm^2\cdot sr} 
 \cdot\left(\frac{2}{k^2\mu^2  \ln\left(\frac{\mu}{m}\right) }\right)\\
 \times \Bigg [k(k^2+2mk-2m^2)\ln\bigg(\frac{(2k-m)(\mu +m-k)}{mk}\bigg) \nn \\
 - \frac{3}{2}k^3 - (\mu+5m)k^2 + (\frac{1}{2}\mu^2 +3\mu m + \frac{9}{2}m^2)k - m^2(\mu+m)  \nn \\
 + \frac{k^2 m^2}{\mu +m-k} + (8k^4-8mk^3-\frac{5}{2}m^2k^2+4m^3k-m^4)\frac{k}{(2k-m)^2}\Bigg] \nn
\ee
where we have taken into account that the total number of positroniums
formed is 4 times the number of positroniums in the $^1S_0$ state 
emitting 511 keV photons with the flux
$\frac{d\Phi}{d\Omega}\simeq 0.025 $ photons cm$^{-2}$s$^{-1}$sr$^{-1}$.
We also took into account the normalization
(\ref{normalization}) for the direct annihilation $e^+e^-\rightarrow 2\gamma$ for large $\mu$.
In eq.(\ref{flux}) mass, $m$, photon energy, $k$ and chemical potential, $\mu$ are all
measured in   MeV units.

 This 
normalization allows us to analyze  the MeV spectra without a precise model
of dark matter /visible matter distributions within the galactic bulge, 
assuming of course  that 
both emissions (511 keV line and 1-20 MeV photons)
come from the same source, the antimatter nuggets. 
As we shall see in the next subsection,
our mechanism can easily explain a   large spectral flux measured by
 COMPTEL if the  values 
of $\mu $ within the physically relevant range of tens of ~MeV and 
parameter $\chi\sim 0.1$, see below\footnote{roughly speaking, the parameter $\chi$  describes the  survival rate of electrons after they enter the nugget's electrosphere 
and  experience the resonance positronim formation.}.

\subsection{Discussion of results}

The  COMPTEL observations, figure \ref{spectra}, suggest that a typical 
spectral flux  in the tens of MeV region is, 
$ k^2\frac{d\Phi(k)}{d\Omega d k}\sim 10^{-2} {\rm (MeV\cdot s^{-1}\cdot cm^{-2}\cdot sr^{-1})} $.
Such a magnitude  can be easily accommodated by our mechanism with a survival rate of $\chi\sim 0.1$ 
as can be seen
from the general normalization of eq. (\ref{flux}). We do not attempt in this work to 
make a precise fitting to the measured spectrum. Such a fitting would require, for example,
the subtraction of all contributions from background emission processes, such as 
those due to cosmic rays\cite{Strong}, or other background astrophysical processes, see 
\cite{MeV} and references to the original literature therein.

On the theoretical side, it is expected that the spectrum may 
be considerably  changed when the surface details of the nuggets 
are taken into account. This is due to the fact that $\mu(r)$ depends 
on the distance from the nugget's surface. 
In this ``transition region" the lepton chemical potential slowly
interpolates between $\mu$ in the bulk and   zero  in the vacuum \cite{olinto}.
The implications of this transition region for the predicted spectrum are
discussed briefly in the appendix below.

However, we do not expect the general normalization factor  $\chi$ as estimated above
to experience any considerable   changes when all these (and many other) unaccounted effects
are considered.

   As argued in \cite{Zhitnitsky:2006tu}, 
    a typical time scale for a formation of the  positronium   is simply a
    typical atomic time if an electron is surrounded by positrons with atomic densities.
 In our case  one expects $\tau_{Ps}\sim \nu^{-1}$ where $\hbar \nu\simeq m_e\alpha^2$
 is a typical   energy scale for the positronium.   Therefore, the vast majority of
 incident electrons will form
 positronium.
 Only a very small 
 portion of the electrons may be expected to avoid  positronium formation, 
 and reach surface where the local chemical potential is large and  
 $e^+e^-\rightarrow 2\gamma$ annihilation events producing 1-20 MeV photons
 dominate. Precise estimation
 of the fraction $\chi$ of electrons which are able to reach the surface strongly 
 depends on a number of factors, 
 such as  typical velocities of electrons in the galactic bulge, specific 
 features of the transition region
 (which itself  depends on  $\mu$ in the bulk, the temperature of the surface 
 and many other parameters).
 It is not our goal of this paper
 to estimate the parameter  $\chi$ using some model dependent calculations, instead
 our goal is to  constrain the properties of the antimatter nuggets 
 using  the observational data available. We are quite satisfied with the 
 result $\chi\ll 1$ which is a natural value  for all types of antimatter nuggets  
 with any type of color superconductivity  in the bulk. In the Appendix we sketch 
 the procedure of calculating the spectrum by using a very simplified model 
 for the nugget's structure. The results of a representative calculation are presented on Fig.1
 where we use numerical coefficient $\chi\sim 0.2$.
 
\section{ Conclusion.}
The discussions in this paper have been motivated by the  observation that the  
  511 keV line and excess of the diffuse $\gamma$ -rays in 1-20 MeV range---two apparently
unrelated puzzles of modern astrophysics---might in fact have a common origin.
It is remarkable that both of these observations can be naturally explained within a  
model which was invented to  explain a completely different puzzle---the
similarities between dark matter  and visible baryonic matter densities in the universe,
$\Omega_{DM}\sim \Omega_B$, rather than having been designed to specifically explain
either 511 keV or 1-20 MeV $\gamma$-ray emisions \footnote{
It is also quite remarkable that another naively unrelated puzzle of modern astrophysics,
diffuse X ray emission, may also find its natural explanation within the same 
model \cite{FZ}.}. In this respect our proposal is very different from a large 
number of other suggestions  which were specifically invented
to  resolve 511 keV puzzle, see
 \cite{Pospelov:2007xh} for an almost complete set of references on the subject. 

Another motivation is to bring to the attention of the astrophysics
community the possibility that these two independent observations may
have a common origin.  Evidence for similar morphologies might already
be present in the existing data as discussed in the introduction-- 511 keV emission as well as
excess of the continuum 1-20 MeV emission (after subtracting contributions from well known astrophysical processes)  
are   concentrated around the galactic center while no considerable excess flux
is observed outside this region. 

If such a morphological  correlation 
between the 511 keV line and the excess of the diffuse $\gamma$-rays at 1-20 MeV
is confirmed by future, more precise measurements, it would give a strong
evidence that the diffuse $\gamma$-rays in the 1-20 MeV range are due to $e^+e^-$
annihilation (511 keV line obviously is a result of $e^+e^-$
annihilation via the positronium formation). At the same time, the required  energetic 
positrons produced as a result of the annihilation of $\sim $ 20 ~MeV 
dark matter particles (suggested, e.g. in\cite{Boehm:2003bt},\cite{MeV-1}) seem ruled out 
 \cite{Beacom:2005qv, Zhang:2006fr,Yuksel:2006fj}. The only option which remains
is to lock energetic positrons $\sim $ 20 MeV in some form for which the in-flight annihilation
with electrons from the interstellar medium is limited to satisfy the necessary
constraints  \cite{Beacom:2005qv,Yuksel:2006fj} (see however ref.\cite{Cembranos:2007fj}
with another suggestion).
 Our model with antimatter nuggets
offers precisely this kind of structure for the confinment of positrons while still allowing
energetic annihilations when electrons from  the interstellar medium  hit the nuggets.
It is quite remarkable that typical values of the estimated lepton chemical potential 
for quark matter fall in the range of tens of MeV which is precisely where an excess  
of diffuse $\gamma$-rays is observed by COMPTEL. We have to stress again, 
we have not introduced this parameter in order to explain the COMPTEL $\gamma$ excess, rather
this range of $\mu\sim $ tens MeV was calculated long ago for a quark matter surface\cite{olinto}.
We should also remark here that very unusual behavior of the spectrum in this region, see Fig.1
is in fact a consequence of some generic features of the nugget's properties
such as presence of electro sphere,  see detail discussions in  Appendix.

Finally, one should notice that there has been a number of 
attempts to explain the  same puzzle ( the excess of the diffuse $\gamma$-rays at 1-20 MeV) using e.g. decaying dark matter particles. However, most models based on this idea already in contradiction with observations, see
recent preprint \cite{Yuksel:2007dr}.
 
  If our testable prediction of spatial correlations between the 511 keV
and  the excess of the diffuse $\gamma$ -ray emission  in 1-20 MeV range
   is verified by new observational data, this would
be of fundamental physical interest, irrespective of any model
specific details.  It would unambiguously imply that
the positrons are hidden in some form of antimatter nuggets.
The point of this paper is to argue that this model
should be seriously investigated because such a correlation might
unlock several important cosmological and astrophysical mysteries.

  \section*{Acknowledgements}

We are  thankful to Andrew Strong, Hasan Yuksel and Xuelei Chen for very useful comments.
We also thankful to  Michael  Forbes for the discussions.
 This work was supported in part by the National Science and Engineering
Research Council of Canada. 

\appendix
\section{  Dark  Matter Anti-Nuggets and their Interactions with In-falling  Visible Electrons}

In order to present a representative spectrum resulting from 
the interactions between a dark matter CCO and incident 
galactic electrons it will be 
necessary to detail some features of the transition region discussed above.
While a full treatment is beyond the scope of this paper some simple calculations
should suffice to demonstrate the general effects of including a non-trivial distribution
of lepton chemical potentials.

While the quark matter surface is relatively sharp, with a scale set by strong force interactions, 
the surrounding leptons are bound electromagnetically 
and take the form of an extended electrosphere. In 
what follows we shall denote the lepton chemical potential at
the quark matter surface as $ \mu_0$, it is this value 
which is expected to 
fall in the $ \sim $ tens  MeV  range. Above the surface the 
local value of $\mu$ must fall off such 
that  $\mu(r\rightarrow\infty)\sim 1/r$ as a consequence of Maxwell's equations and the
requirement of chemical equilibrium \cite{olinto}. For present purposes 
we consider the ultra relativistic case  in which the local positron chemical potential 
is much larger than $m$  
 and varies with distance $z$ from the
 surface  as 
 \be
 \label{mu}
 \mu_{e^+}(z)=\sqrt{\frac{3\pi}{2\alpha}}\frac{1}{(z+z_0)}, ~~~~z_0=\sqrt{\frac{3\pi}{2\alpha}}\frac{1}{\mu_0}, ~~~ n_{e^+}(z)\simeq \frac{\mu_{e^+}^3(z)}{3\pi^2}
 \ee
 where 
$ z_0    $ can be interpreted as a characteristic thickness 
of the electrosphere \cite{olinto},\cite{Cheng}. Such a behavior (\ref{mu}) is a result of 
mean field calculations similar to the Thomas-Fermi approximation in atomic physics. 
This model has previously been applied in the context
of quark stars\cite{olinto},\cite{Cheng}.

At a given height the flux of incident electrons will undergo an exponential 
extinction with an annihilation rate $ \frac{dI(z)}{dt} $ as given 
in equation (\ref{normalization}). Consequently the number of electrons surviving
at height z   satisfies, 
\be
\label{density}
\left( \frac{dN_{e-}(z)}{dt} \right)  = N_{e-}(z) \left( \frac{dI(z)}{dt} \right)  
\ee
In principle, if we knew the precise structure of the 
electrosphere at all heights, from small to very large densities,   we 
could calculate the survival rate $\chi$ as well as the  number  of electrons 
$ N_{e-}(z)$ available for annihilation with the nugget's positrons in the dense region closest
to the quark matter surface.
  As we mentioned above,
a full treatment of this problem is beyond the scope of the present work. Instead, we 
introduce a phenomenological parameter $\chi$ which describes the fraction of electrons which survive
resonance formation and reach the dense inner region where direct annihilation events
dominate.  We   model the dense  region
by introducing a parameter $z_{max}$ such the relevant region satisfies 
the following condition, $z\leq z_{max}$. We will choose 
$z_{max} $ such that $\mu(z_{max}) \gg m_e $ thus  the ultra relativistic 
approximation holds within this region and eq. (\ref{mu}) can be trusted. 

For large $\mu_{e^+}(z)\gg m$ one can approximately integrate equation (\ref{density}) 
using the expression (\ref{mu}) for  $\mu_{e^+(z)}$ and approximate expression (\ref{normalization}) for 
$ \left( \frac{dI(z)}{dt} \right)  $ at large $\mu$. The obtained result is
 \be
\label{density1}
N_{e-}(z)   = N_{e-}(z_{max}) e^{-\Bigg[\frac{3\alpha}{4m v_e}\left(\frac{1}{(z+z_0)}-\frac{1}{(z_{max}+z_0)}\right) \ln\left(\frac{\mu_0}{m}\right) \Bigg]}, ~~ z\leq z_{max},
\ee
where $v_e$ is the $z$- component of the in-falling electron's velocity.  
In order to obtain the complete resultant spectrum 
we average expression (\ref{flux}), calculated
based on annihilation at a specific $ \mu_{e^+}(z) $ value,
over all heights weighted by the remaining electron density at point $z$ above the nugget's surface,
\be
 \frac{d\Phi(k)_{net} }{d\Omega d k}\sim \int^{z_{max}}_{0} dz N_{e-}(z)  \frac{d\Phi(k, \mu(z))}{d\Omega d k}.
\ee
As is to be expected accounting for the distribution of $ \mu(z) $ values with height $z$
produces an $ E^2 I $ curve which is considerably flatter than the single
$ \mu $ valued spectrum given in (\ref{flux}). The result of this procedure
is shown in Fig \ref{spectra} with representative values of $\chi = 0.2, \mu_0=60, z_{max}=30z_0 $ chosen.

We stress that we do not attempt 
in the present work to make a perfect fit to the observations
by exploring a variety of possible models for the electrospheres of the nuggets.
Instead, the main goal of this appendix is to present an idea of how a  calculation of 
the  complete spectrum resulting
from our mechanism would proceed in a simple-minded model. We have not attempted to 
consider the effects of a range of incident electron velocities, and have limited our calculations
to the relativistic limit so that the resultant spectrum is unreliable below  $E \sim$ few MeV.
  However, the generic  features of the spectrum will remain the same.
    These features are:
  from eq. (\ref{flux}) it is clear that for a given $\mu(z)$  
the flux times $k^2$ is strongly peaked at  maximum possible  momenta 
$k\leq \mu(z)$. At the same time 
electron density $ N_{e-}(z)  $ is decreasing when electron moves to the 
region of the larger positron densities as eq. (\ref{density1}) suggests.  These 
two effects strongly compensate for each other such
that $E^2\Phi$  is almost flat in the region of interests, varying by only a factor of two over
an order of magnitude in energy. Detailed analysis 
of the spectrum as well as its model dependence  and sensitivity to phenomenological parameters 
will be presented in our future work. 
In particular, for photon energy $k\sim (20-30) $MeV the excess almost vanishes according
to analysis \cite{Strong} while the spectrum above this region could be nicely fitted. Our model automatically 
produces emission in this range as a generic consequence of the exponential suppression of the 
  electron density $ N_{e-}(z)$ when electrons are approaching the very dense region with large $\mu$ close to the  nugget's surface, see eq.  (\ref{density1}). Indeed, superposition of the $\gamma$ ray background calculated in \cite{Strong1} (blue solid line in \ref{spectra}) and the $\gamma$ ray  contribution due to the   annihilation mechanism with nuggets (heavy black dots) falls almost 
  to the central value of the  COMPTEL data at the highest energy about 20 MeV (green vertical bar 
  at the   right side). As we mentioned above, in the few MeV region our calculations are not valid due to
  some assumptions  we have made to simplify our estimates such as cut off at $z_{max}$ where chemical potential $\mu( z_{max}) $  is in few MeV range. Therefore, we interpret the obtained results as   consistent with the COMPTEL data.

\end{document}